\documentclass[useAMS,usenatbib]{mn2e}
\usepackage{graphicx,epsfig}

\title{High-redshift X-ray properties of the haloes of simulated disc 
galaxies}

\author[J. Rasmussen et al.]
{J.~Rasmussen$^1$\thanks{E-mail: jr@astro.ku.dk},
J.~Sommer-Larsen$^2$, S.~Toft$^1$, K.~Pedersen$^1$ \\
$^1$ Astronomical Observatory, University of Copenhagen, 
Juliane Maries Vej 30, DK-2100 Copenhagen \O, Denmark\\
$^2$ Theoretical Astrophysics Center,
Juliane Maries Vej 30, DK-2100 Copenhagen \O, Denmark}

\date{}

\pagerange{\pageref{firstpage}--\pageref{lastpage}} \pubyear{2003}
\def\LaTeX{L\kern-.36em\raise.3ex\hbox{a}\kern-.15em
T\kern-.1667em\lower.7ex\hbox{E}\kern-.125emX}

\begin{document}
\label{firstpage}
\maketitle

\begin{abstract}
X-ray luminosities and surface brightness profiles of the hot gas haloes of
simulated disc galaxies at redshifts $z=0-2$ are presented. The galaxies
are extracted from fully cosmological simulations and correspond in mass
to the Milky Way. We find that the bolometric X-ray luminosities of the haloes
decrease by a factor $4-10$ from $z\sim 1$ to $z\sim 0$, 
reflecting the decrease in the rate at which hot halo gas cools out on to the
disc. At all redshifts, most of the emission is found to originate within 
10--15 kpc of the disc.
When combined with models in which the evolution
of disc X-ray luminosity is dominated by X-ray binaries, the predicted halo
luminosities at $z\sim 1$ show good agreement with constraints from spiral
galaxies in {\it Chandra} Deep Field data.
There is an indication that haloes with a metal abundance of 0.3Z$_{\odot}$ 
overpredict observed X-ray luminosities at $z \sim 1$, suggesting that 
halo metallicities are lower than this value. Prospects for direct detection 
of the haloes of Milky Way--sized galaxies with current and future X-ray
instrumentation are discussed. It is found that {\it XEUS} should be able to
single out the halo emission of highly inclined Milky Way--sized disc 
galaxies out to $z \approx 0.3$. For such galaxies in this redshift interval,
we estimate a lower limit to the surface density of detectable haloes on the 
sky of $\sim 10$ deg$^{-2}$.
More generally,
owing to their luminosity evolution, the optimum redshifts at which to
observe such haloes could be $0.5<z<1$, depending on their assembly history.

\end{abstract}


\section{Introduction}\label{sec,intro}

Standard models of disc galaxy formation require spiral galaxies to be 
surrounded by large reservoirs of hot gas which should be emitting at X-ray 
wavelengths and from which gas should still be accreting on to the disc at
present (e.g.\ \citealt{whi91}).
The detailed properties of these haloes and their role in the 
formation and evolution of galactic discs remain largely unknown from an
observational viewpoint, as the haloes have so far escaped direct detection.
Possible exceptions are NGC891 (\citealt{bre97}; \citealt{str03a}) 
and the Milky Way itself (e.g.\ \citealt{sid96}; \citealt{pie98}), 
if neglecting cases where (i) the 
galactic discs show evidence of being disturbed by tidal interactions, or (ii)
the X-ray emission at off-disc distances of a few kpc can be attributed to 
processes originating in the disc such as feedback from star formation
(\citealt{dah03}; Strickland et~al.\ 2003a,b). A possible explanation for this
lack of halo detections may be provided by recent simulation work 
(\citealt{bir03}; \citealt{kat03}), which
suggests that most galaxies accrete the majority of their gas at temperatures
much lower than the virial temperature of their halo. This would be 
particularly pronounced at higher redshifts, and would imply that most of the
halo radiation is emitted as Ly$\alpha$ emission close to the disc rather 
than as soft X-rays from an extended, quasi-spherical region.

In a previous paper (\citealt{to}, hereafter Paper I) we extracted
disc galaxies from fully cosmological hydrodynamical simulations 
and determined their present-day halo X-ray luminosities, finding 
consistency with observational upper limits on halo emission from nearby 
spirals. It should be mentioned that we have subsequently detected an error
in the X-ray calculations employed in that study. Correcting this increases 
the luminosity of hot haloes with gas metallicity $Z= 0.0$ by a factor
$\sim 2$ and that of $Z\simeq 0.3$Z$_{\odot}$ haloes by a factor $\sim 1.25$.
This correction applies equally to all our simulated galaxies, so the overall 
conclusions of Paper I remain unchanged. Accounting for this error still
produces results in consistency with observations, as the upper limits on
observed luminosities can easily accommodate the correction.
One main result of the study of Paper~I was the conjecture that spiral haloes
were
possibly one order of magnitude brighter in soft X-rays at $z=1$ than today,
a result which was based solely on the predicted mass accretion rates of the 
discs.
Analytical models of halo emission overpredict 
emission at $z=0$ by at least an order of magnitude (see \citealt{ben00} and
Paper~I) and thus cannot be assumed to provide a reliable description of the
evolution in X-ray emission with redshift.
To amend this situation, we here extend our previous work by studying the 
high-redshift properties of a few disc galaxies extracted from cosmological
cold dark matter simulations.

We found in Paper~I that the halo luminosity approximately scales as 
$L_X \propto V_c^5$, where $V_c$ is the characteristic circular speed 
in the disc at $R_{2.2} = 2.2 R_d$ and $R_d$ is the disc scalelength.
Although subject to considerable scatter, such a trend is consistent with 
expectations from simple cooling flow models and suggests that massive 
galaxies ($V_c \ga 300$ km s$^{-1}$) would be optimum targets for 
observing halo emission. 
Such galaxies are rare, however, and hence do not compare well to the typical
spiral seen in deep X-ray surveys. The aim of this work is therefore to study 
the predicted properties of, and the detection prospects for, haloes of more 
typical disc galaxies of size similar to the Milky Way (MW).

The simulations and X-ray computations are described in \S \ref{sec,sim}. 
Results are presented in \S \ref{sec,res} and compared to observations in 
\S \ref{sec,obs}. We discuss the possibility for detecting the haloes in 
\S \ref{sec,det} and summarize our findings in \S \ref{sec,sum}.

\section{Simulations and X-ray calculations}\label{sec,sim}

The primary goal is to study the redshift evolution in halo X-ray 
properties of simulated disc galaxies. To this end, we have extracted two 
disc galaxies from fully cosmological
{\sc Tree}SPH simulations of galaxy formation in a $\Lambda$CDM cosmology with
$\Omega_M=0.3$, $\Omega_{\Lambda}=0.7$, and $H_0=65$ km s$^{-1}$ Mpc$^{-1}$.
These values are adopted throughout this paper.
The simulations include star formation, stellar feedback processes, and a 
meta-galactic UV radiation field. For a 
description of initial conditions and simulation details, we refer to
\citet*{som03}.

From a simulation with an adopted universal baryon fraction of $f_b = 0.10$,
we study in detail two galaxies (hereafter denoted {\tt gal15}
and {\tt gal18}), each run both with a cooling function based on a primordial 
gas composition ([Fe/H]$=-\infty$, i.e.\ $Z=0.0$) and with one based on the 
typical intracluster gas metallicity 
([Fe/H]$=-0.5$, i.e.\ $Z\simeq 0.3$Z$_{\odot}$). The latter abundance can 
probably {\it a priori} be considered a reasonable upper limit to the 
metallicity of hot halo gas.
The two galaxies were selected from the requirement that they should
show disc circular speeds $V_c$ at $z=0$ comparable to that of the 
present-day MW. With values at $z=0$ of $V_c=225$ (250) km s$^{-1}$ for 
{\tt gal15} with
$Z=0.0$ ($Z=0.3$Z$_{\odot}$) and $V_c=202$ (213) km s$^{-1}$ for {\tt gal18} 
with $Z=0.0$ ($Z=0.3$Z$_{\odot}$), they have MW-like masses and hence 
correspond to $L\sim L^{\ast}$ 
galaxies in the local Universe. The $z=0$ version of both galaxies were
included in Paper I.
Since the galaxies were formed in the same cosmological simulation,
differences between them only reflect their different assembly 
histories.

For the calculations of halo X-ray properties we follow the approach of 
Paper I, to which we refer for more details: All SPH gas particles outside a 
box sized (1000 kpc)$^3$ centred on the galaxy are cut away, as is 'cold' gas 
(with $T < 3\times 10^4$ K). Spatially smoothed temperature and density 
fields are 
constructed from the values associated with individual particles, and X-ray 
volume emissivities are calculated from these using the {\sc meka} plasma 
emissivity code \citep*{me}.

At any given redshift, some variation in total X-ray luminosity is seen on
time-scales of a few tens of Myr, due to 
galaxy-galaxy interactions and merging of smaller satellites. 
Since we are mainly interested in the long-term evolution of $L_X$,
on the order of 10 frames separated by
time intervals $\Delta t\approx 100$ Myr have been extracted around each 
redshift.
This time interval is large enough that individual frames can be considered
reasonably independent, given the typical time-scale involved, yet small
enough for overlaps between adjacent redshift bins to be insignificant.
X-ray properties of each frame have been calculated as prescribed above,
and frames have then been omitted for which the total bolometric 
(0.012--12.4 keV) luminosity deviates by more than $2\sigma$ from the mean at
that redshift (discarding on average less than one frame per redshift, 
corresponding to $\sim 7$ per cent of all frames).
Mean values and dispersions of all quantities of interest were derived at a
given redshift from the remaining frames and used in the subsequent analysis.

For the $Z=0.0$ galaxies we follow the halo properties back to
$z=2.9$ ({\tt gal15}) and $z=2.3$ ({\tt gal18}), beyond which the disc itself 
is not well-defined. The 
$Z=0.3$Z$_{\odot}$ galaxies show a more complex gas distribution 
at all redshifts and are followed only back to $z\approx 2$.

\section{Results}\label{sec,res}

\subsection{$L_X$ versus redshift}

Of prime interest is the evolution with redshift of the X-ray 
luminosity, in part because this is expected to reflect the gas accretion 
history of the stellar disc as will be discussed below.

Fig.~\ref{fig,lxz} shows the redshift evolution of the rest-frame 
0.2--2 keV and 
bolometric luminosities $L_{X,bol}$ of the hot haloes of the two galaxies. 
In general $L_{X,bol}$ is seen to increase with redshift, rising fairly 
steeply out to $z\sim 1$, beyond which
the increase levels off. The increase from $z=0-1$ is a factor $\sim 4-10$, 
thus verifying our prediction from Paper I that disc galaxy haloes could be
up to an order of magnitude brighter at $z=1$ than present-day haloes.
As can be seen, {\tt gal15} shows a steady evolution in both X-ray bands, 
whereas {\tt gal18} exhibits a slightly more complex behaviour. 
The reason is that {\tt gal18} experiences a period of enhanced accretion,
as will be demonstrated below.

\begin{figure*}
\begin{center}
\mbox{\hspace{-0.5cm}
\epsfxsize=9.2cm
\epsfysize=7cm
\epsfbox{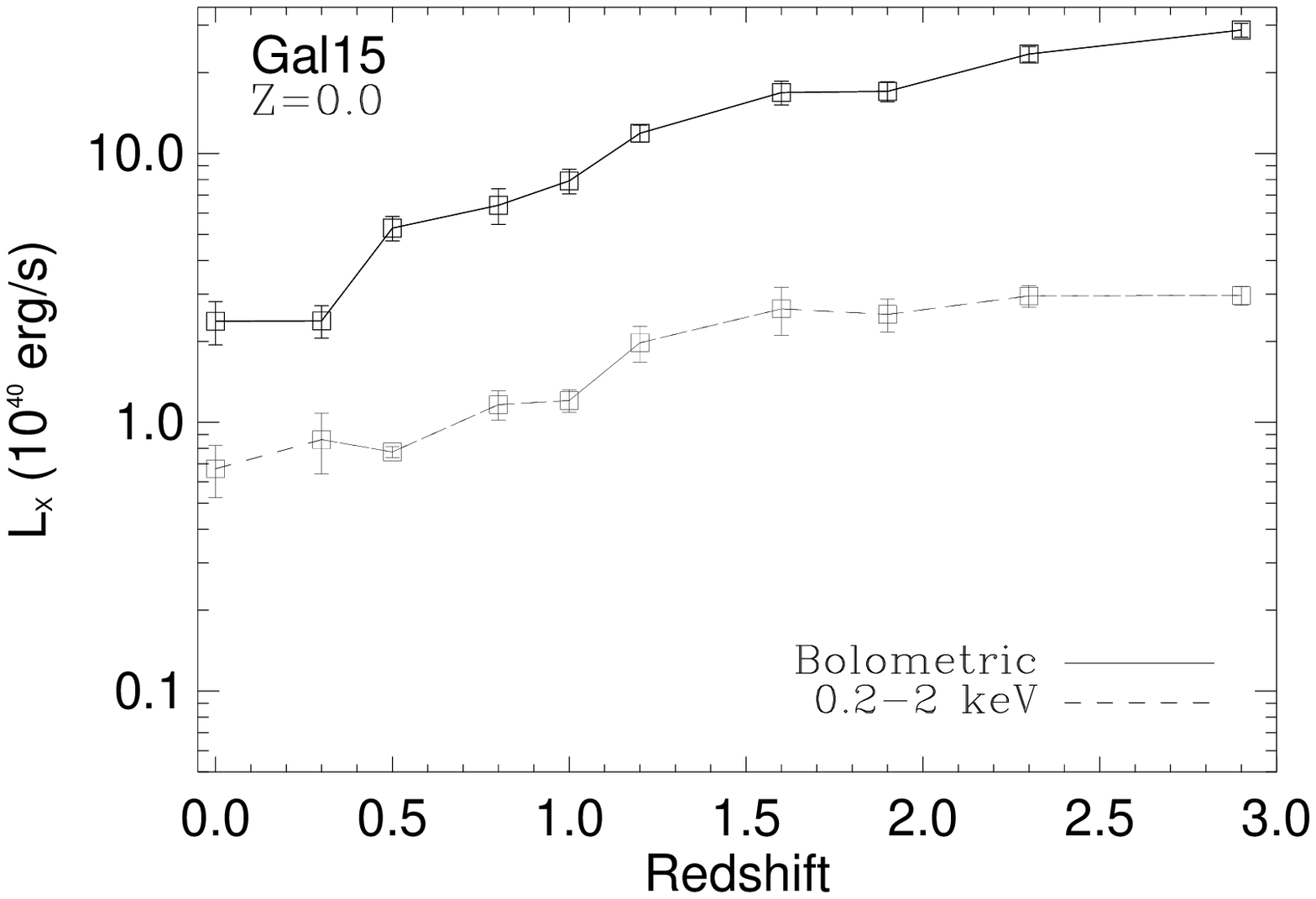}\hspace{0cm}
\epsfxsize=9.2cm
\epsfysize=7cm
\epsfbox{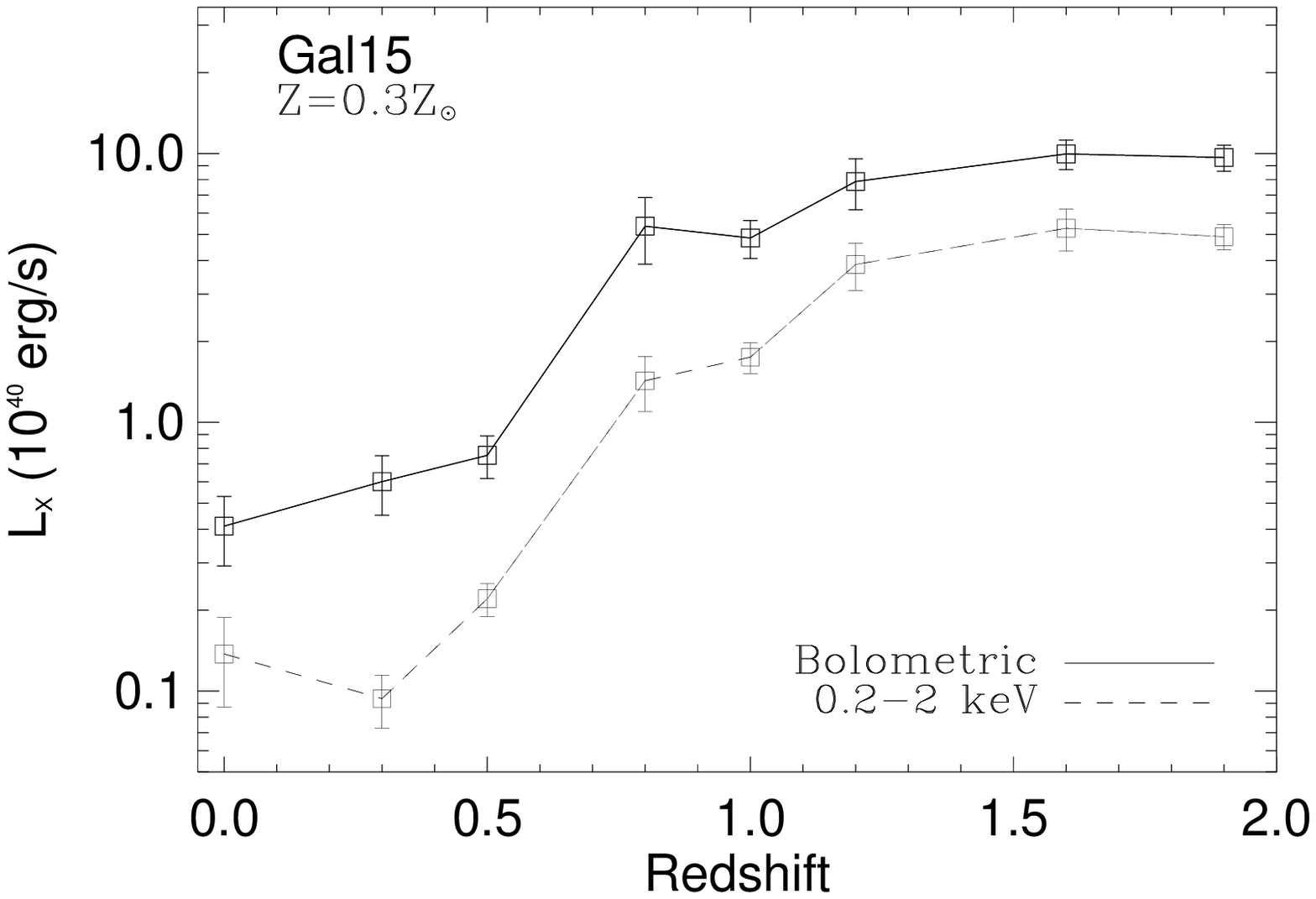}\hspace{0cm}}
\mbox{\hspace{-0.5cm}
\epsfxsize=9.2cm
\epsfysize=7cm
\epsfbox{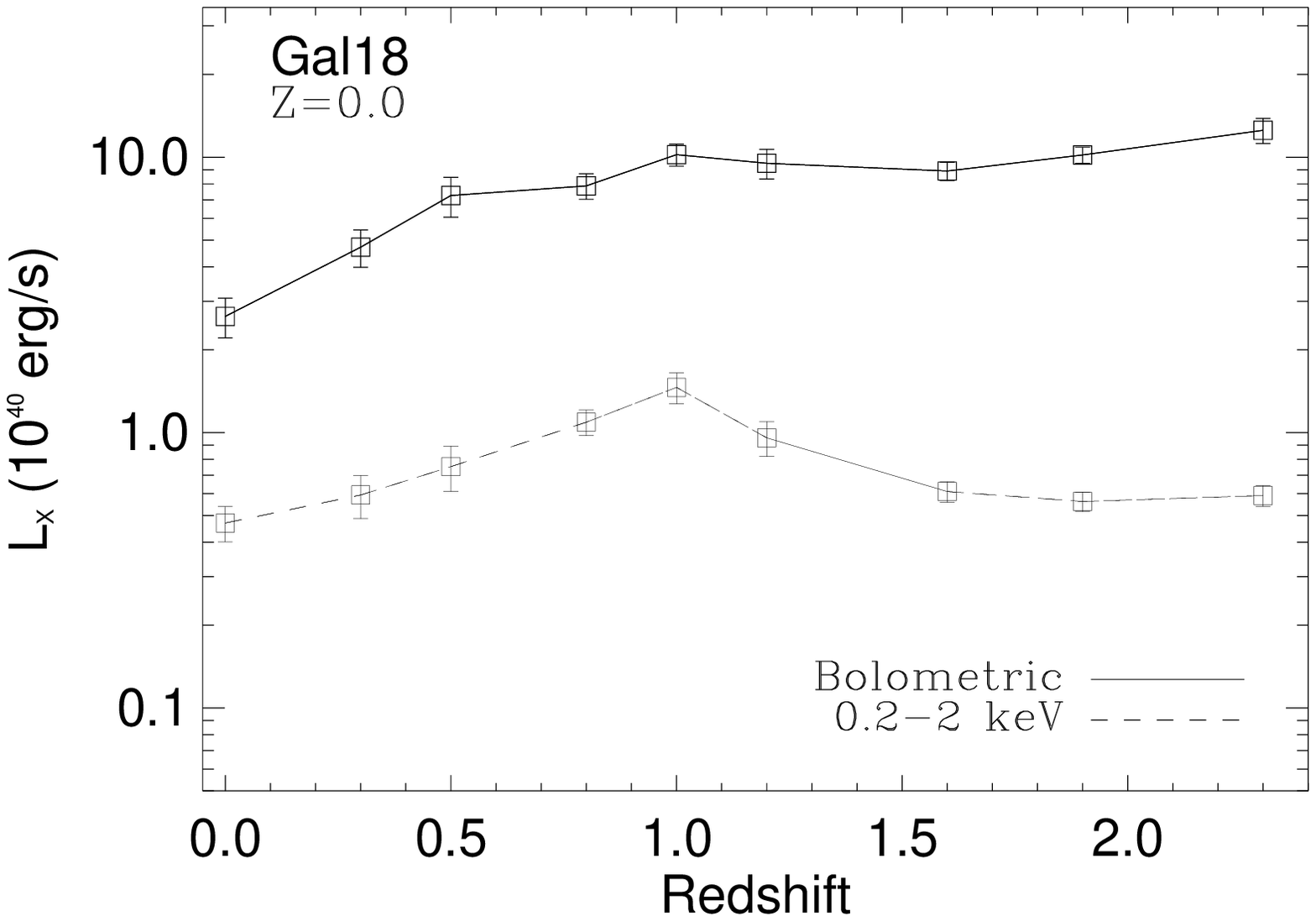}\hspace{0cm}
\epsfxsize=9.2cm
\epsfysize=7cm
\epsfbox{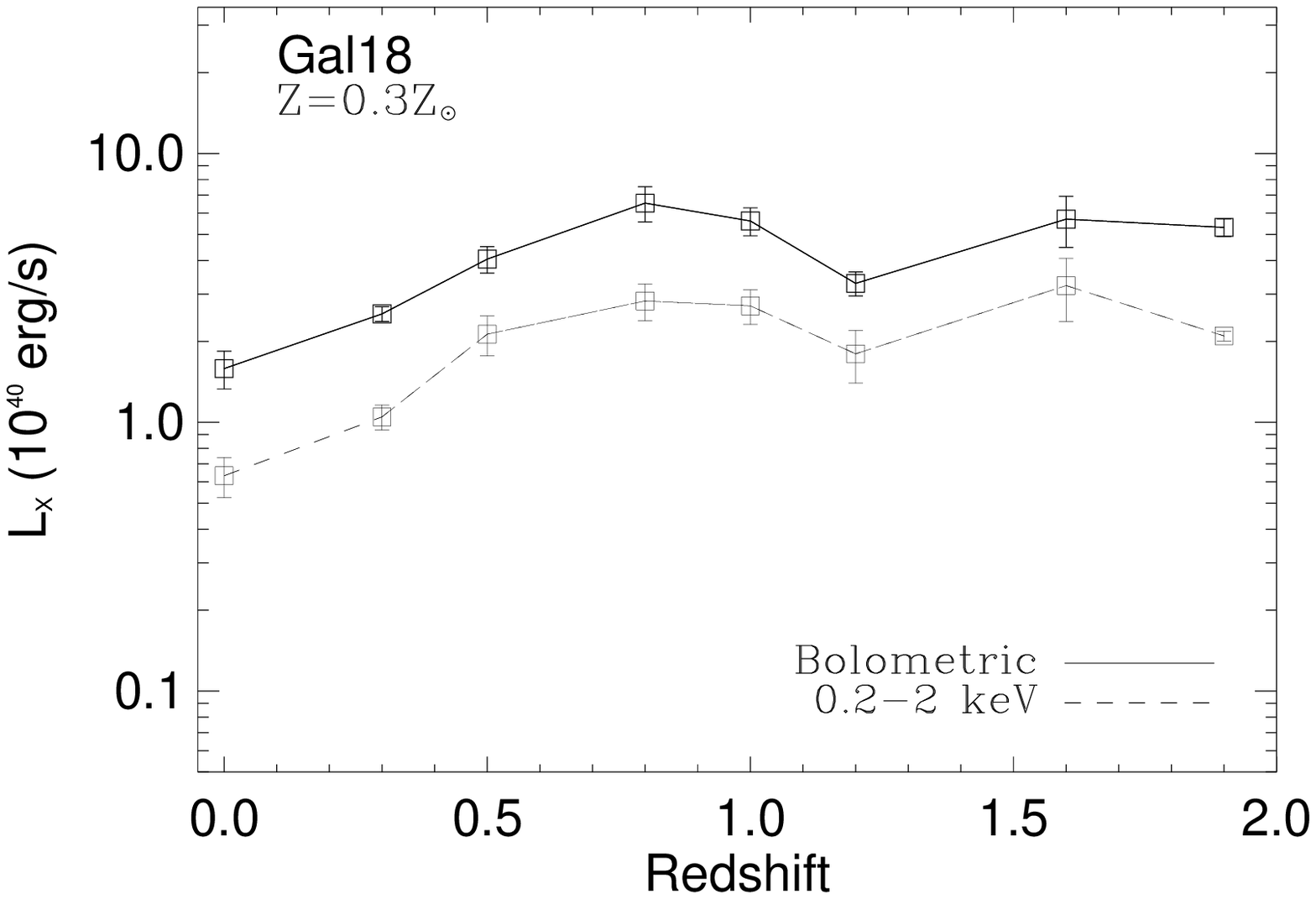}\hspace{0.0cm}}
\end{center}
\caption{Redshift evolution of the rest-frame bolometric and 0.2--2 keV X-ray
luminosities of the simulated galaxies. 
Top panel {\tt gal15}, bottom panel {\tt gal18}, 
with metallicities as labelled. Error bars signify the $1\sigma$ dispersion
between frames at a given redshift.}
\label{fig,lxz}
\end{figure*}

It is seen that at high $z$ the 0.2--2 keV emission is somewhat higher for 
the $Z=0.3$Z$_{\odot}$ hot haloes than for the $Z=0.0$ haloes, falling short 
of the bolometric 
luminosity at all $z$ only by a factor of $\sim 2-3$. 
This is because the $Z=0.3$Z$_{\odot}$ haloes display higher 
emission-weighted mean temperatures
($\sim$~0.20--0.25 keV, with little redshift dependence) than the $Z=0.0$
haloes ($\sim$~0.10--0.15 keV), thus
radiating a relatively larger fraction of their total emission in this band.
Given the near-constancy of the hot gas temperature, the decrease in 
$L_X$ with time is predominantly due to a decline in the available amount of 
hot halo gas along with a decrease in the volume-weighted mean density of 
this gas. The mass of hot halo gas in the X-ray emitting volume declines 
only by a factor $\sim 2$ ($Z=0.0$) and $\sim 4$ ($Z=0.3$Z$_{\odot}$)
from $z\sim 1$ to $z\sim 0$, however.
An implication of this combined with Fig.~\ref{fig,lxz} is that
haloes radiate relatively more per unit mass of hot gas at high $z$ than at 
$z=0$ (typically by a factor $\sim 2$ at $z=1$ relative to $z=0$).

Note that for both galaxies, increasing the cooling efficiency by 
raising the gas metallicity to $Z=0.3$Z$_{\odot}$ leads to a decrease in 
bolometric $L_X$ at $z=0$, consistent with the results from our much larger
 sample in Paper I. The decrease is less significant for {\tt gal18}, which, 
as will be argued below, is 
probably somewhat atypical of the galaxies studied in that paper.
Also note that this decrease persists back to at least $z\approx 2$. This may
at first seem counter-intuitive, given the higher X-ray emissivity of a 
$Z=0.3$Z$_{\odot}$ plasma.
The reason is that already at redshifts $z\ga 2$ a larger fraction of hot 
halo gas
has cooled out to form stars in the $Z =0.3$Z$_{\odot}$ galaxies, leading at 
$z\approx 2$ to
a factor $\sim 2$ deficit in the total amount and volume-weighted mean
density of hot gas relative to the $Z=0.0$ case.

\subsection{Accretion rates}

The halo luminosity can be linked to the rate at 
which hot halo gas cools out and accretes on to the disc, as illustrated by 
the following considerations.

The cooling time is the gas energy density divided by the cooling rate,
\begin{equation}
t_{cool} \approx  \frac{\frac{3}{2} NkT}{\Lambda n_H^2},
\end{equation}
where $\Lambda$ is the cooling function, $n_H$ is the hydrogen number density,
and $N=(n_e +\sum_i n_i)$ is the total number density of electrons and all 
ion species $i$. 
The mass cool--out rate d$\dot M$ of gas with mass density $\rho$ in a 
small volume element d$V$ becomes
\begin{equation}
\mbox{d}\dot M \la \frac{\rho \mbox{d}V}{t_{cool}} = \frac{2\mu m_p}{3kT} 
\Lambda n_H^2 \mbox{d}V,
\end{equation}
where $\mu$ is the mean molecular weight, $m_p$ the proton mass, and
equality corresponds to the case of absence of external heat sources and 
$p\mbox{d}V$ work.
Since the emission-weighted mean inverse temperature is
\begin{equation}
\langle 1/T \rangle_{ew} = \frac{ \int \Lambda n_H^2 \frac{1}{T} \mbox{d}V}{ \int \Lambda n_H^2 \mbox{d}V} 
\end{equation}
and the bolometric luminosity is $L_X = \int \Lambda n_H^2 \mbox{d}V$, 
we find
\begin{equation}\label{eq,acc}
\dot M \la \frac{2\mu m_p}{3k} \int_V \frac{\Lambda n_H^2}{T} \mbox{d}V 
= \frac{2\mu m_p}{3k} L_X \langle 1/T\rangle_{ew}      
\end{equation}
$$ \approx 0.066 \left(\frac{L_X}{10^{40} \mbox{ erg s$^{-1}$}} \right) \left(\frac{\langle1/T\rangle_{ew}}{\mbox{keV$^{-1}$}}\right) \mbox{ M$_{\odot}$ yr$^{-1}$}.$$
In this simplified picture, we therefore expect $\dot M$ to (roughly) scale 
with $L_X\langle 1/T\rangle_{ew}$. 

As was shown in Paper I, the gas in the simulations 
is essentially two-phased, so $\dot M$ is here estimated at any given 
redshift by considering the rate at which hot gas 
($T\ga 2\times 10^6$ K) cools out to the cold phase ($T\la 3\times 10^4$ K), 
see \citet{som03}.

In Fig.~\ref{fig,acc} we have plotted $\dot M$ derived for the $Z=0.0$ 
galaxies along with the expectation from Eq.~(\ref{eq,acc}). There is a 
remarkably good agreement in shape, particularly for the steadily 
accreting galaxy {\tt gal15}. Although the agreement is slightly less 
convincing for 
{\tt gal18}, it is, however, clear that this galaxy undergoes a period of 
strongly enhanced accretion around $z\approx 1.3$ which affects its X-ray
luminosity accordingly. 
In both cases the accretion rate at
$z\approx 1$ is found to be $\sim 5$ times larger than at present, decreasing
e.g.\ 
for {\tt gal15} from 2.5 to 0.5 M$_{\odot}$ yr$^{-1}$ from $z=1$ to $z=0$. 
Based on Fig.~\ref{fig,acc} (left), an approximate relation
describing this evolution in the redshift range $z=0-1$ is
\begin{equation}
\mbox{log}\dot M(z) \approx 0.6z - 0.25 ,
\end{equation}
with $\dot M$ in M$_{\odot}$ yr$^{-1}$. The observed difference in 
normalization between this relation and Eq.~(\ref{eq,acc}) is due to the 
simplifying assumptions underlying the latter.
We speculate that the main reasons for the discrepancy are in part the 
neglect of $p\mbox{d}V$ work and in part a geometrical effect related to
the fact that cooling proceeds in a cooling flow, so considering a fixed 
volume for the estimation of $\dot M$ is a clear simplification. The problem
is highly complex, however, and we cannot exclude that other effects play a 
role too.

\begin{figure*}
\mbox{\hspace{-0.8cm}
\epsfxsize=9.3cm
\epsfysize=7cm
\epsfbox{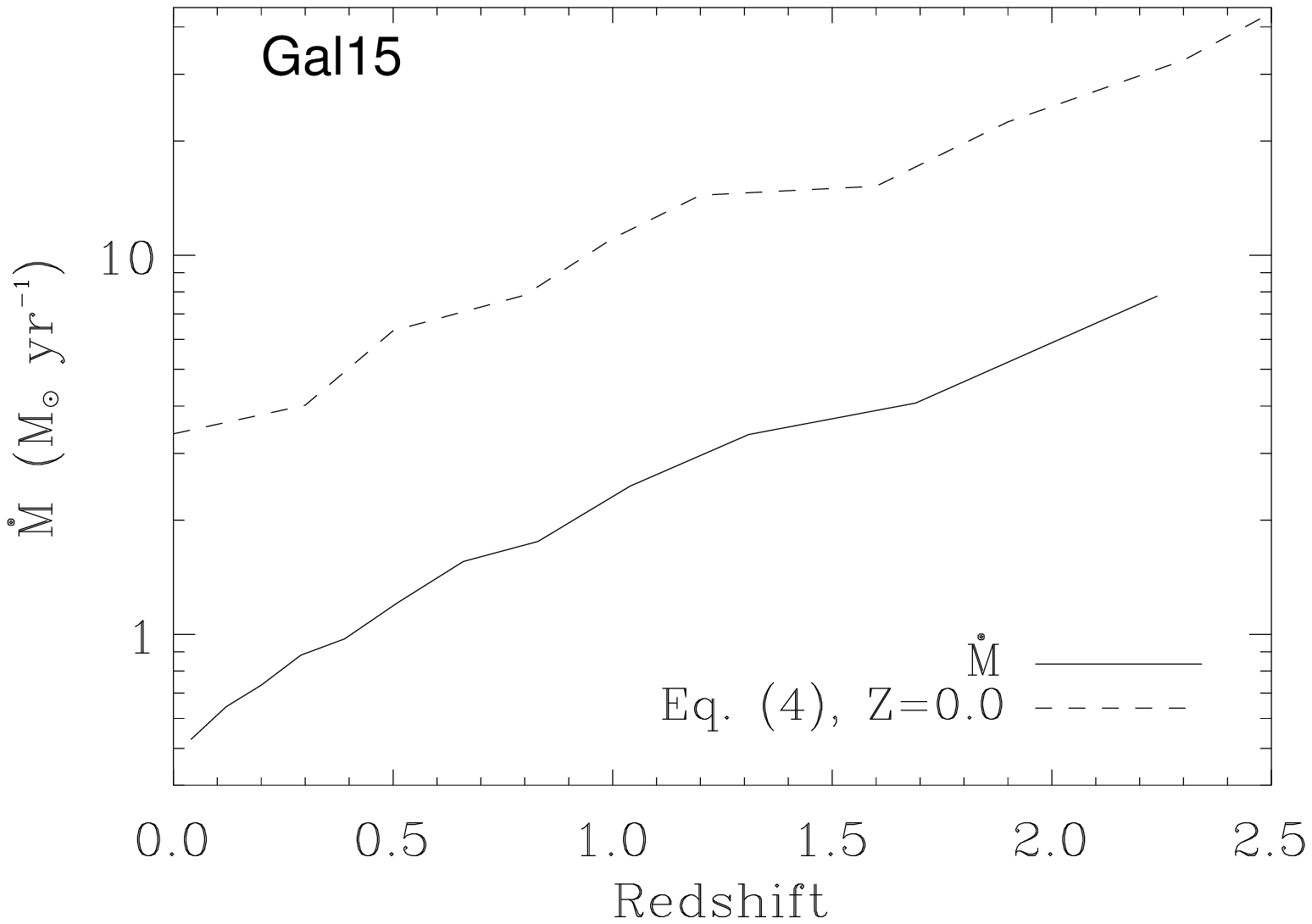}\hspace{0cm}
\epsfxsize=9.3cm
\epsfysize=7cm
\epsfbox{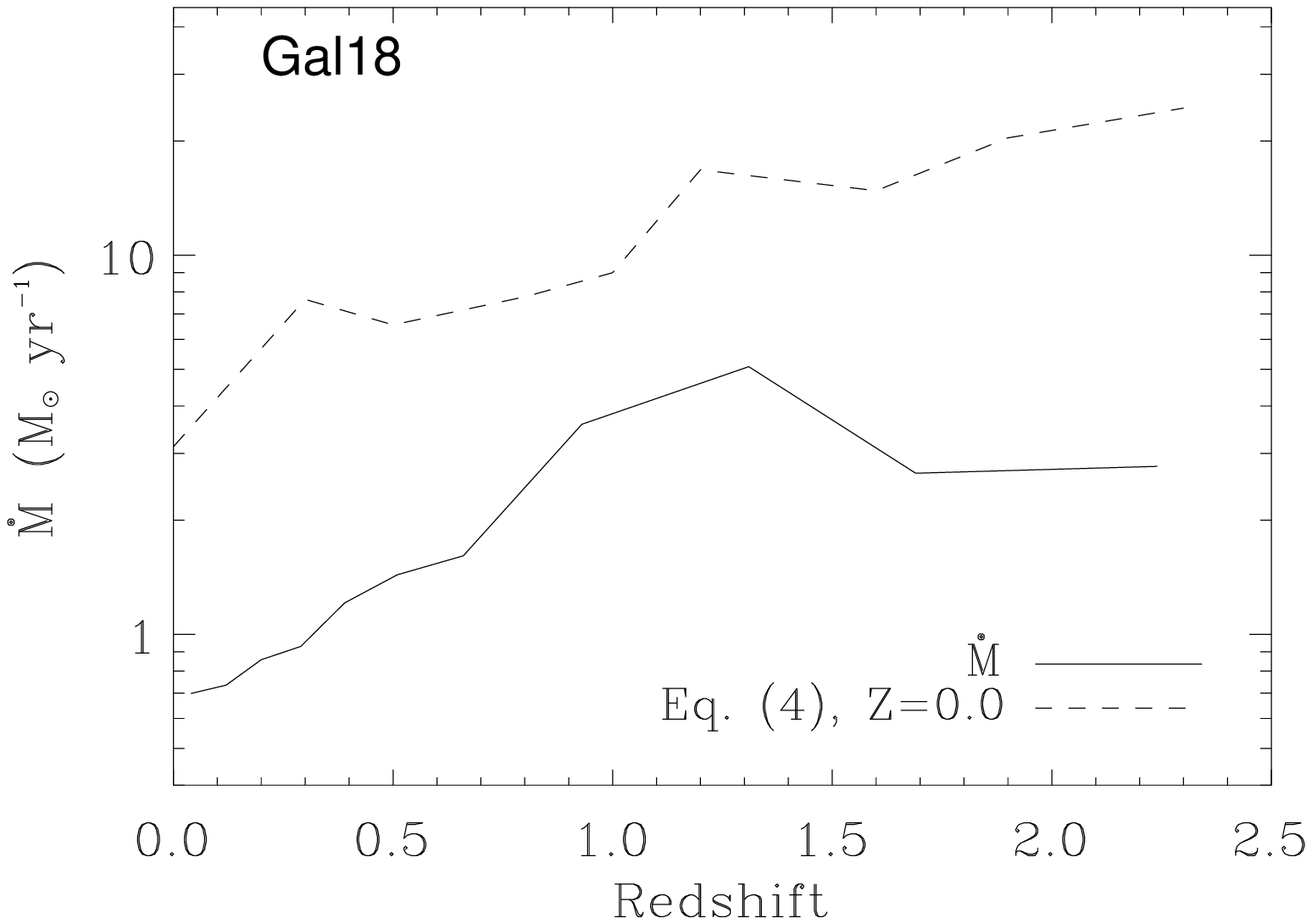}\hspace{0cm}}
\caption{Total disc accretion rate $\dot M$ (solid line) of gas, and the 
quantity $L_{X,bol}\langle 1/T\rangle$, normalized according to 
Eq.~(\ref{eq,acc}).} 
\label{fig,acc}
\end{figure*}

Given the predicted evolution of disc accretion rates and halo bolometric 
X-ray luminosities, it is interesting to note that \citet{bar01} found that
accretion on to supermassive black holes, as measured by the bolometric
X-ray luminosity of active galactic nuclei (AGN), follows a very similar 
trend to that seen in Figs.~\ref{fig,lxz} and
\ref{fig,acc}. The logarithm of the black hole mass accretion rate 
was found to grow roughly linearly with redshift out to $z\sim 1$--1.5, 
above which the increase levels off.
It seems reasonable to assume that a positive correlation should exist 
between the disc accretion rate and that of any central black hole
(given the observed correlation between black hole mass and host galaxy bulge
mass, e.g.\ \citealt{mcl02}), in which case the 
observations of \citet{bar01} could be taken to support the overall trends 
predicted in Figs.~\ref{fig,lxz} and \ref{fig,acc}.

\section{Comparison with observations}\label{sec,obs}

Although only a few relevant observational constraints on halo emission are 
available even in the nearby Universe,
we showed in Paper I that $L_X$ of our simulated  galaxies at $z=0$ agree 
with observed upper limits for haloes of nearby massive spirals and with 
estimates of the amount and luminosity of hot halo gas in the Milky Way
(these conclusions all remain valid after correction for the error 
mentioned in \S \ref{sec,intro}).
Some recent observational progress in this context include the results
of \citet{kun03} on emission in the nearby
spiral M101. These authors find that our halo prediction from Paper I 
for a galaxy like M101 ($V_c \approx 170$ km s$^{-1}$) can easily be 
accommodated within the amount of diffuse emission obtained for this galaxy, 
provided the halo gas density displays a sufficiently small vertical scale
height.
Other recent results comprise those of \citet{dah03}, Strickland et~al.\
(2003a,b), and \citet*{wan03}, but in these cases the X-ray emission is 
either affected by 
tidal interaction with another galaxy or can be attributed to processes in 
the disc, and they cannot be directly compared to our models.

Although at high redshifts there is currently no possibility of comparing the 
predicted levels
of halo emission to direct observations, useful constraints 
can nevertheless be extracted from the {\it Chandra} Deep Field observations 
(these are superior in sensitivity to any other pointed X-ray observation).
Spiral galaxies, classified morphologically from 
{\it Hubble} Deep Field data, have been detected in the 
1 Ms {\it Chandra} Deep Field-North (CDF-N) data down to a nominal limiting 
0.5--2 keV point source flux of 
$\approx 3\times 10^{-17}$ erg cm$^{-2}$ s$^{-1}$, and using a stacking 
technique detection has been performed to even lower fluxes \citep{hor02}. 
The latter authors considered the integrated X-ray emission 
from $L \approx L^{\ast}$ Sa--Sc spirals at $z=0.4-1.5$ 
which did not harbour a bright AGN or display strong 
starburst activity. Resulting rest-frame 0.5--2 keV 
luminosities in several redshift ranges were derived and can be taken here
as observational upper limits to our predicted halo emission.
Accounting for disc emission, not included in our simulated galaxies, 
strengthens the CDF constraints on halo emission further. To this end,
we consider the MW disc, for which \citet{war02} lists a 
total 0.5--2 keV luminosity of $\sim 5\times 10^{39}$ erg s$^{-1}$, including
all stellar and diffuse disc sources. This provides a
measure of the contribution to be added to that of our simulated 
haloes at $z=0$, in order to obtain an estimate of the integrated X-ray
luminosity of the simulated galaxies. 
However, the MW disc is dominated by X-ray binaries (XRBs) at a level of 
$\sim 3\times 10^{39}$ erg s$^{-1}$ \citep{war02}. This contribution to the
disc emission is 
unlikely to remain constant with redshift, as it is expected
to roughly scale with the disc star formation rate (SFR). Globally, the 
latter is known
to rise a factor of $\sim 10$ between $z\sim$ 0--1 (e.g.\ \citealt*{mad98}),
a result which also applies to the CDF-N in particular \citep{coh03}. 
Based on an assumed such coupling between disc SFR and $L_X$ of the XRB 
population, \cite{gho01} present a model for the redshift evolution of $L_X$ 
from XRBs. 
Assuming a Madau-type SFR profile, these authors predict a factor of 
$\sim 5$ (3) increase in $L_X$ from XRBs at $z=1$ ($z=2$) 
relative to $z=0$. 
We have adopted the expected XRB contribution to the integrated
disc emission from \citet{war02} and evolved it with redshift according to
the \citet{gho01} model. An addition from other disc and bulge X-ray sources
of $2\times 10^{39}$ erg s$^{-1}$, assumed constant for simplicity, is 
included in order to match the observed total MW disc luminosity.
The resulting disc 
luminosity of an MW-like galaxy is compared in Fig.~\ref{fig,cdfn} (left) 
to the nominal limits in different redshift bins from the CDF-N sub-samples 
of galaxies with
spectroscopically confirmed and purely photometric redshifts, respectively.
Typical uncertainties on the observed X-ray luminosities are displayed 
for comparison; the $1\sigma$ statistical errors are
$\sim$20 per cent (A.\ Hornschemeier, priv.\ comm.), very similar at the 
relevant redshifts to those of our simulated galaxies.
Adopting the mean $L_X$ in the sample of \citet*{sha01} as representative
of the typical spiral $L_X$ at $z=0$ (comparable to the MW value adopted 
here), \citet{hor02} note that the observed increase in $L_X$ from $z=0$ to 
$z=0.6$ is consistent with the prediction of the 
\citet{gho01} model. When taking the CDF values as upper limits, 
Fig.~\ref{fig,cdfn} suggests that this consistency can be 
extended to redshifts $z \approx 1.5$, provided that $L_X$ of the other disc
X-ray emitting components, such as hot diffuse gas in the disc and bulge, 
does not increase dramatically between $z\approx 0.5$ and $\approx 1.5$. 
Some increase in
diffuse disc luminosity is probably expected, however, due to supernova driven
feedback associated with the increase in star formation with redshift.

\begin{figure*}
\mbox{\hspace{-0.3cm}
\epsfxsize=9.1cm
\epsfysize=7.2cm
\epsfbox{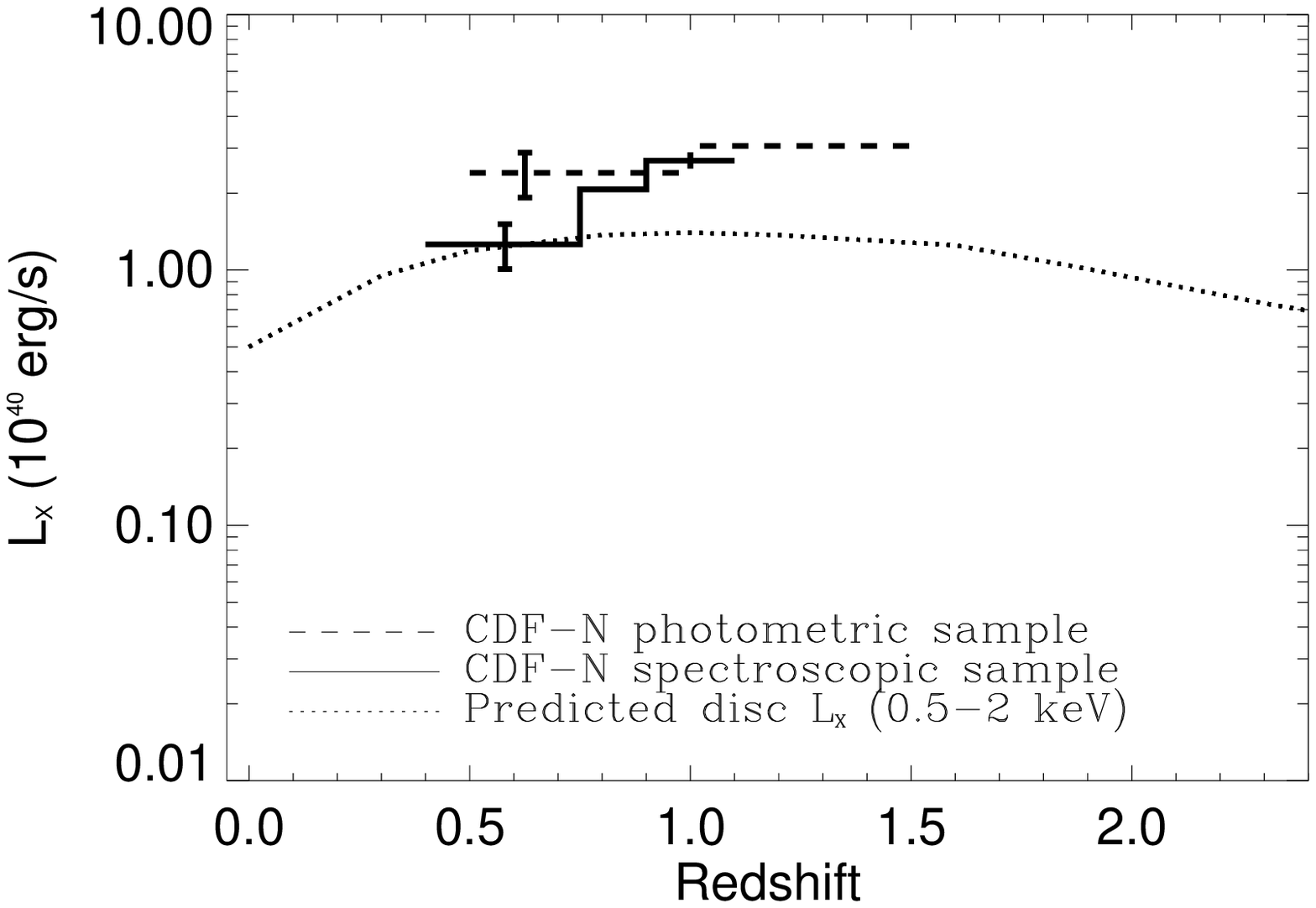}\hspace{0cm}
\epsfxsize=9.1cm
\epsfysize=7.2cm
\epsfbox{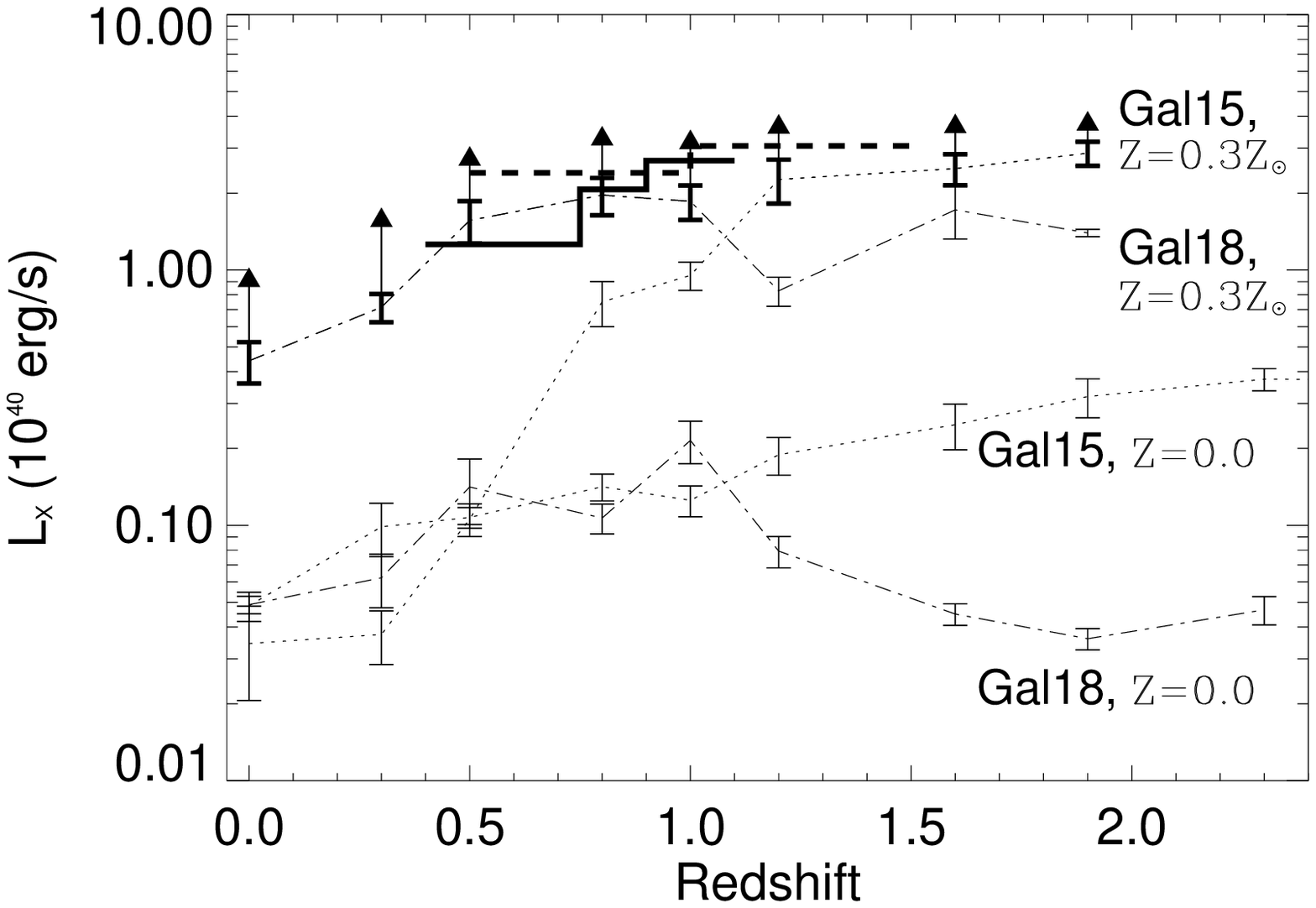}\hspace{0cm}}
\caption{{\it Left}: {\it Chandra} Deep Field-North results (0.5--2 keV) on 
$L\approx L^{\ast}$ galaxies, along with the predicted evolution in 
{\it disc} X-ray luminosity of a MW-like galaxy, assuming that X-ray binaries 
drive the evolution (\citealt{gho01}; \citealt{war02}).
Solid line is the CDF-N subsample of galaxies with spectroscopic redshifts, 
dashed line the subsample with photometric redshifts only, with typical
$1\sigma$ errors overplotted.
{\it Right}: CDF-N results and our predicted intrinsic 0.5-2 keV halo 
luminosities. Dotted line is {\tt gal15}, while dot-dashed line is 
{\tt gal18}, with metallicities as labelled. Upward arrows show the shift in 
$L_X$ resulting from adding the disc luminosities of the left figure.} 
\label{fig,cdfn}
\end{figure*}

\citet{coh03} has recently shown that the CDF-N spirals out to $z=1.4$ follow 
an $L_X$--SFR relationship comparable to that seen for local galaxies,
albeit with larger scatter.
This result supports the approach employed by \citet{gho01} in their XRB 
model. Motivated by this, and by Fig.~\ref{fig,cdfn} (left), we compare in 
Fig.~\ref{fig,cdfn} (right) the CDF constraints to our halo 
predictions, having added the scaled \citet{gho01} model (plotted as upwards 
arrows) to the upper limit of the highest predicted halo luminosity at any 
given redshift. Uncertainties on the CDF values are here omitted for clarity.
As can be seen, the $Z=0.0$ galaxies are easily consistent with the CDF-N 
constraints, 
whereas the metal--rich galaxies are much closer to the observed values. 
When adding the \citet{gho01} model to the predicted halo 
luminosities, the agreement with the CDF-N results is fairly good, our
predictions for the integrated emission then being consistent with results 
from the photometric CDF-N sample. 

There is an indication, however, 
that the $Z=0.3$Z$_{\odot}$ haloes, in particular {\tt gal18}, could be 
slightly too X-ray luminous at the relevant redshifts. 
While this is certainly not a large excess, 
we see at least two possible explanations for this.
Possibly {\tt gal18} is peculiar. It certainly displays a high $L_X$ at $z=0$
for its circular velocity and metallicity when compared to other galaxies of 
the study presented in Paper I, supporting this idea. 
Alternatively, Fig.~\ref{fig,cdfn} may suggest that assuming halo 
metallicities of $Z=0.3$Z$_{\odot}$ at {\it all} redshifts probed here is 
unrealistic, overpredicting the 
resulting X-ray emission at the CDF-N redshifts. The fact that {\tt gal15}, 
which at $z=0$ is very typical of the galaxies studied in Paper I, is also 
close to the CDF constraints at $z>1$ would seem to substantiate 
this possibility. While only indicative, such a result would not be 
surprising; as argued 
by e.g.\ \citet{ren03}, the global metallicity of the Universe at 
$z\approx 3$ is likely to be $\sim 0.1$Z$_{\odot}$ rather than 
$\sim 0.3$Z$_{\odot}$. On the other hand,
for haloes at $z=1$, the predicted unabsorbed 0.5--2 keV flux 
originating inside a region $1\times 1$ arcsec$^2$ on the sky (within the
central $\sim 10\times 10$ kpc of the halo) is
$\sim 2\times 10^{-19}$ and $\sim 6\times 10^{-19}$ erg cm$^{-2}$ s$^{-1}$ 
for the $Z=0.3$Z$_{\odot}$ haloes of {\tt gal15} and {\tt gal18}, 
respectively.
This is well below the CDF-N point source flux for normal spirals at this 
redshift, $\sim (3-6) \times 10^{-18}$ erg cm$^{-2}$ s$^{-1}$ 
\citep{hor02}. 
Note, however, that we have not attempted to quantify any systematic 
uncertainties on the predicted integrated
emission (related to the choice of SFR model, the coupling between $L_X$ and 
SFR, the assumed XRB fraction of disc $L_X$ etc.). For example, it is not
obviously appropriate to use a Madau-type SFR profile to extrapolate the
SFR of individual galaxies to high redshifts. Taking again the MW as an
example and neglecting variations caused by short-term starbursts, evidence
suggests that the MW disc has experienced a roughly constant SFR back to at 
least $z\sim 1$ \citep{roc00}. Assuming a constant SFR, the 
\citet{gho01} model
predicts a steady decline in disc $L_X$ from $z=0$ to $z=2$ of $\sim 30$ per 
cent, in which case our predictions for the $Z=0.3$Z$_{\odot}$ galaxies do
not violate the CDF constraints, but, in fact, show even better agreement
with these values.

We finally note that when adding the \citet{gho01} model to the $Z=0.0$ 
haloes, the integrated emission is well below that of the CDF values and so 
is easily consistent with these when viewed as upper limits.
Predicted and observed luminosities can in this case be reconciled if 
using the \citet{gho01} model along with a different SFR prescription, e.g.\
the 'hierarchical' model of \citet{bla99}.

\section{Prospects for halo detection}\label{sec,det}

In order to assess the possibility of detecting the X-ray 
haloes of MW-sized discs with current and future X-ray instrumentation, 
we have computed X-ray brightness profiles perpendicular to the discs of the 
simulated galaxies. The discs have been oriented edge-on, and
observer-frame surface brightness (energy flux per unit solid angle) and 
flux profiles as a function of off-disc distance have been calculated in
40 kpc wide and 5--8 kpc high slices parallel to the disc.
Spectra were calculated for each SPH particle in the rest-frame band matching
the observer-frame band of interest, and the resulting total surface 
brightness has been diminished by a factor $(1+z)^4$ to account 
for cosmological dimming. Galactic absorption has been incorporated assuming 
an absorbing column density at $z=0$ of $N_H = 3\times 10^{20}$ cm$^{-2}$,  
using the photoelectric absorption cross sections of \citet{mor83}.
Profiles have been calculated in the 0.3--2 keV band, expected to provide the
optimum signal-to-noise (S/N) ratio for {\it Chandra} and {\it XMM-Newton}, 
given the predicted halo temperatures, as well as telescope effective areas 
and background levels. 
Calculations were repeated for the 0.1--3 keV band, which, in this context, 
is likely to be relevant to the next-generation {\it XEUS} mission, 
being well-calibrated down to energies at least as low as 0.1 keV and 
at high energies providing a much higher S/N than e.g.\ {\it XMM-Newton}.

The 0.3--2 keV profiles are shown in Fig.~\ref{fig,surfbright} for the
most optimistic case of $Z=0.3$Z$_{\odot}$ edge-on galaxies. An immediate
conclusion, apart from the apparent fact that resulting fluxes are exceedingly
low, is that at all redshifts the majority of emission originates within the 
innermost 10--15 kpc, as was also found to be the case at $z=0$ (Paper I).
At least two additional inferences from these plots are worth noting:

\begin{figure*}
\mbox{\hspace{-0.3cm}
\epsfxsize=9.1cm
\epsfysize=7cm
\epsfbox{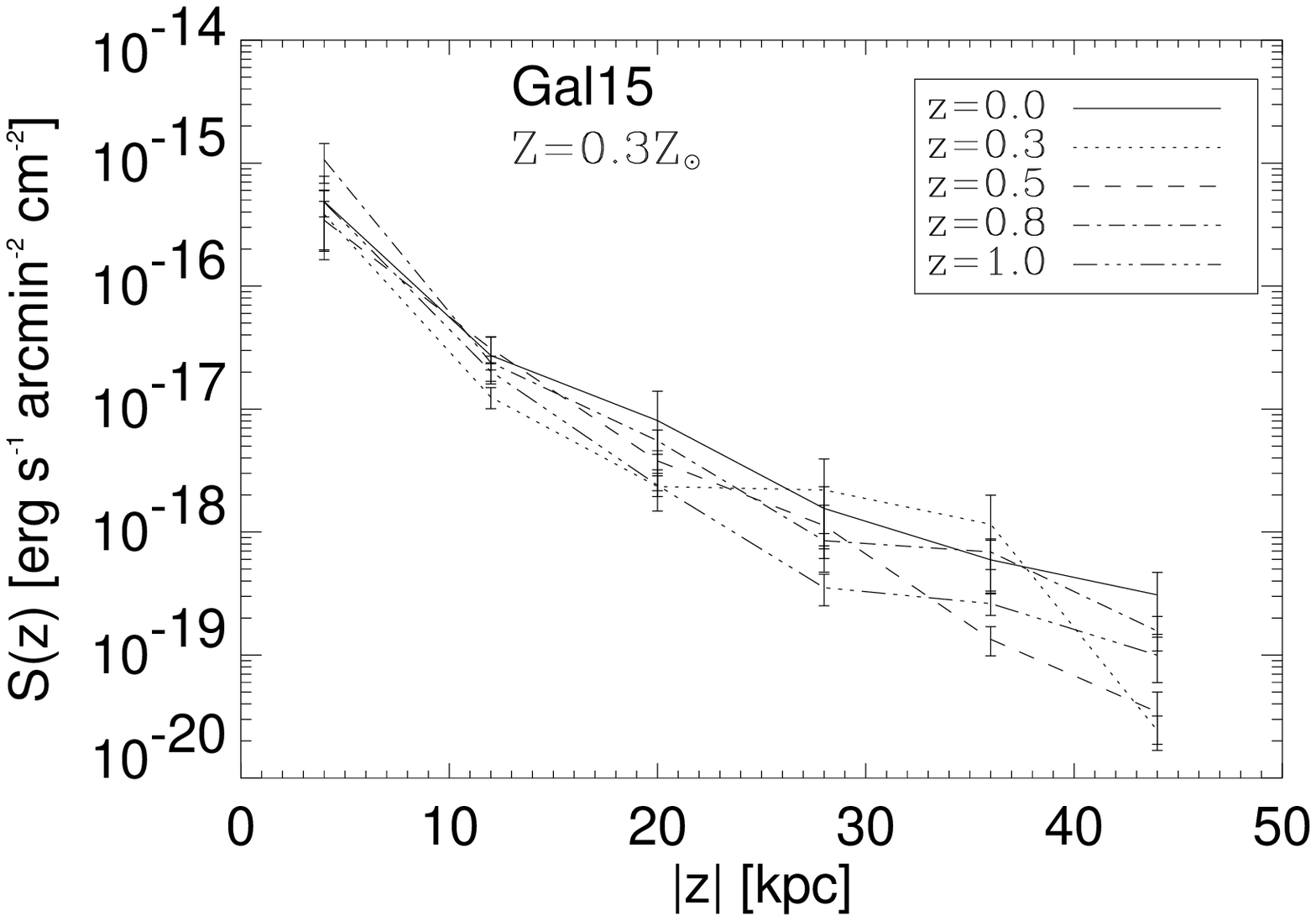}\hspace{0cm}
\epsfxsize=9.1cm
\epsfysize=7cm
\epsfbox{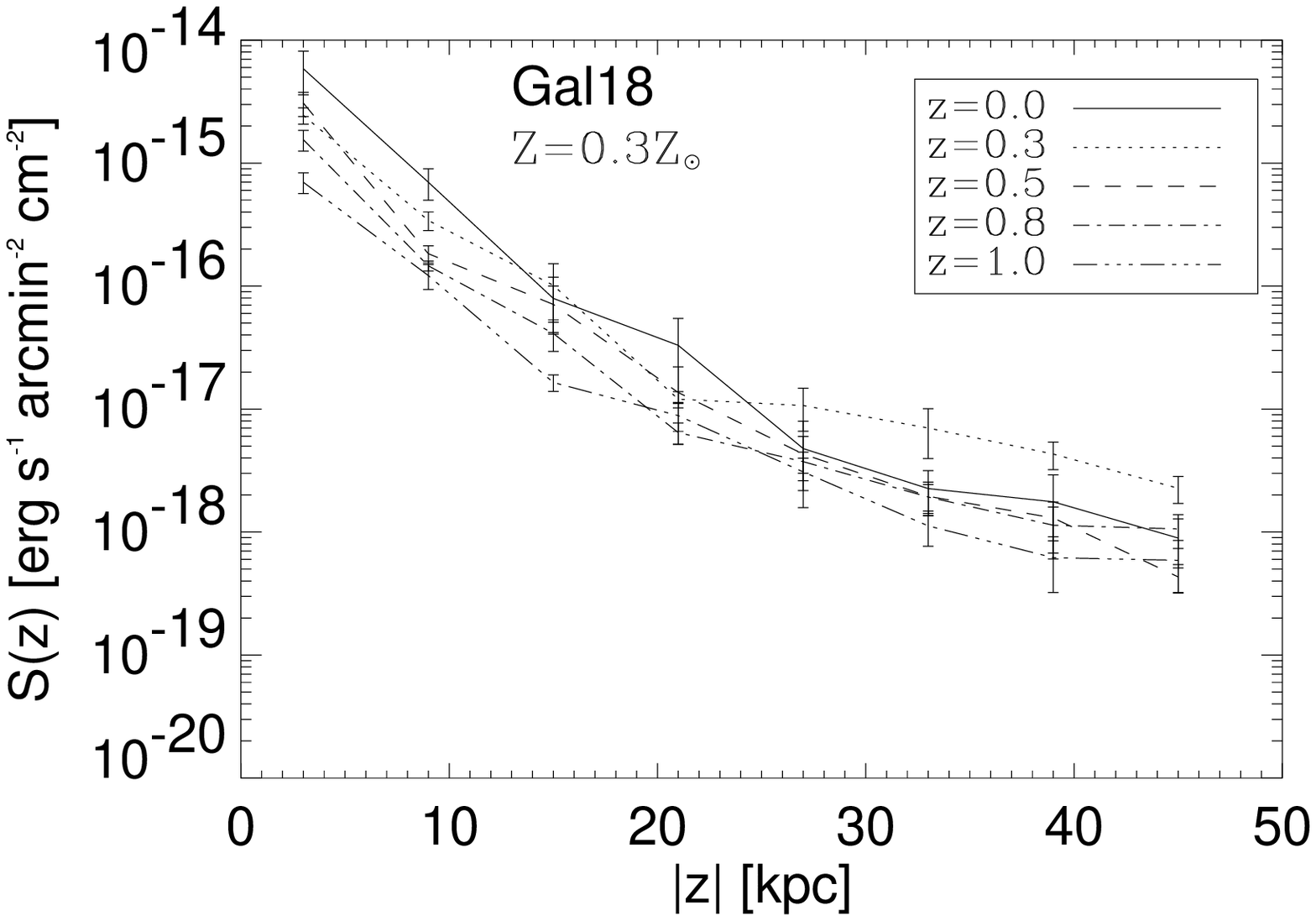}\hspace{0cm}}
\mbox{\hspace{-0.3cm}
\epsfxsize=9.1cm
\epsfysize=7cm
\epsfbox{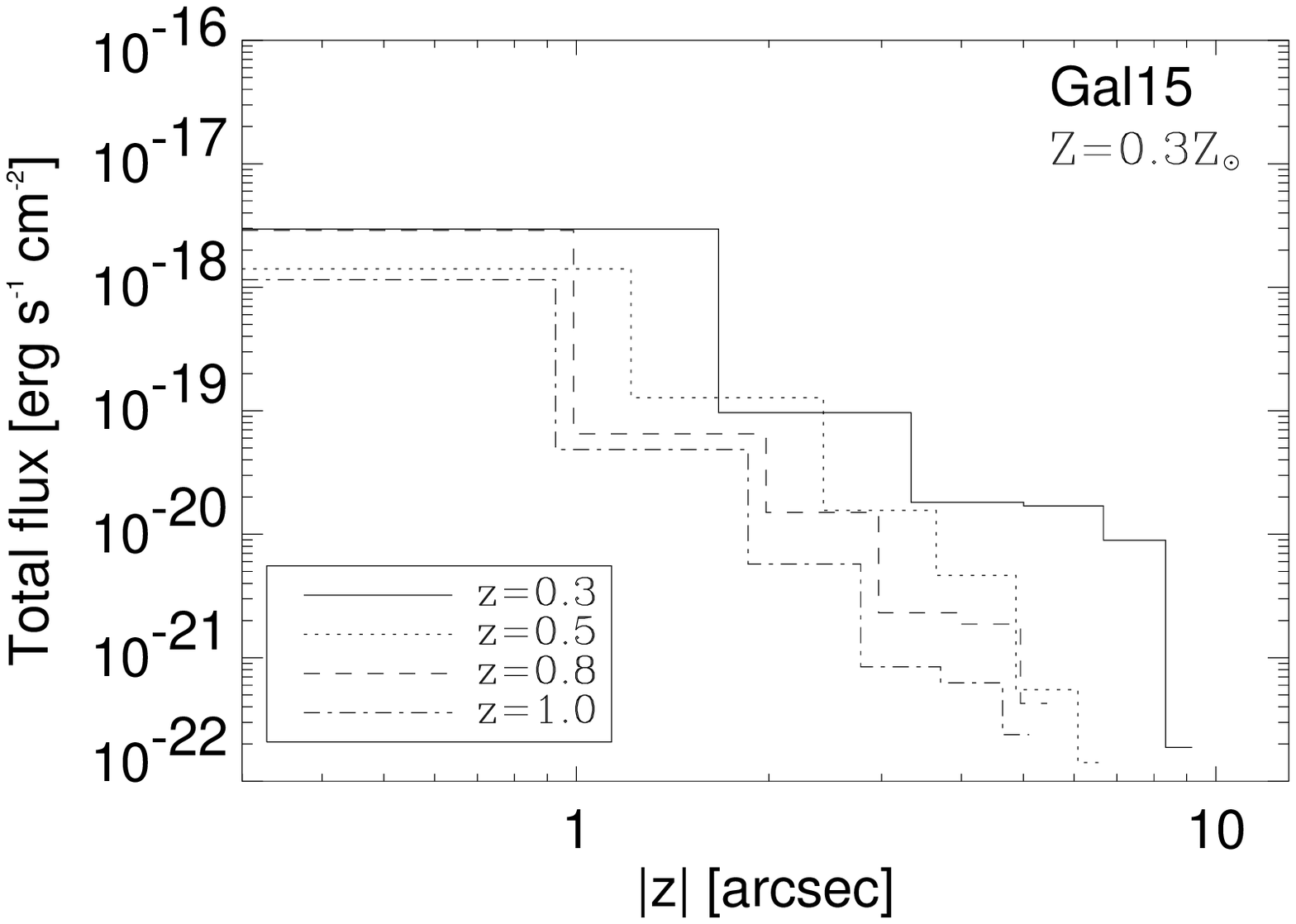}\hspace{0cm}
\epsfxsize=9.1cm
\epsfysize=7cm
\epsfbox{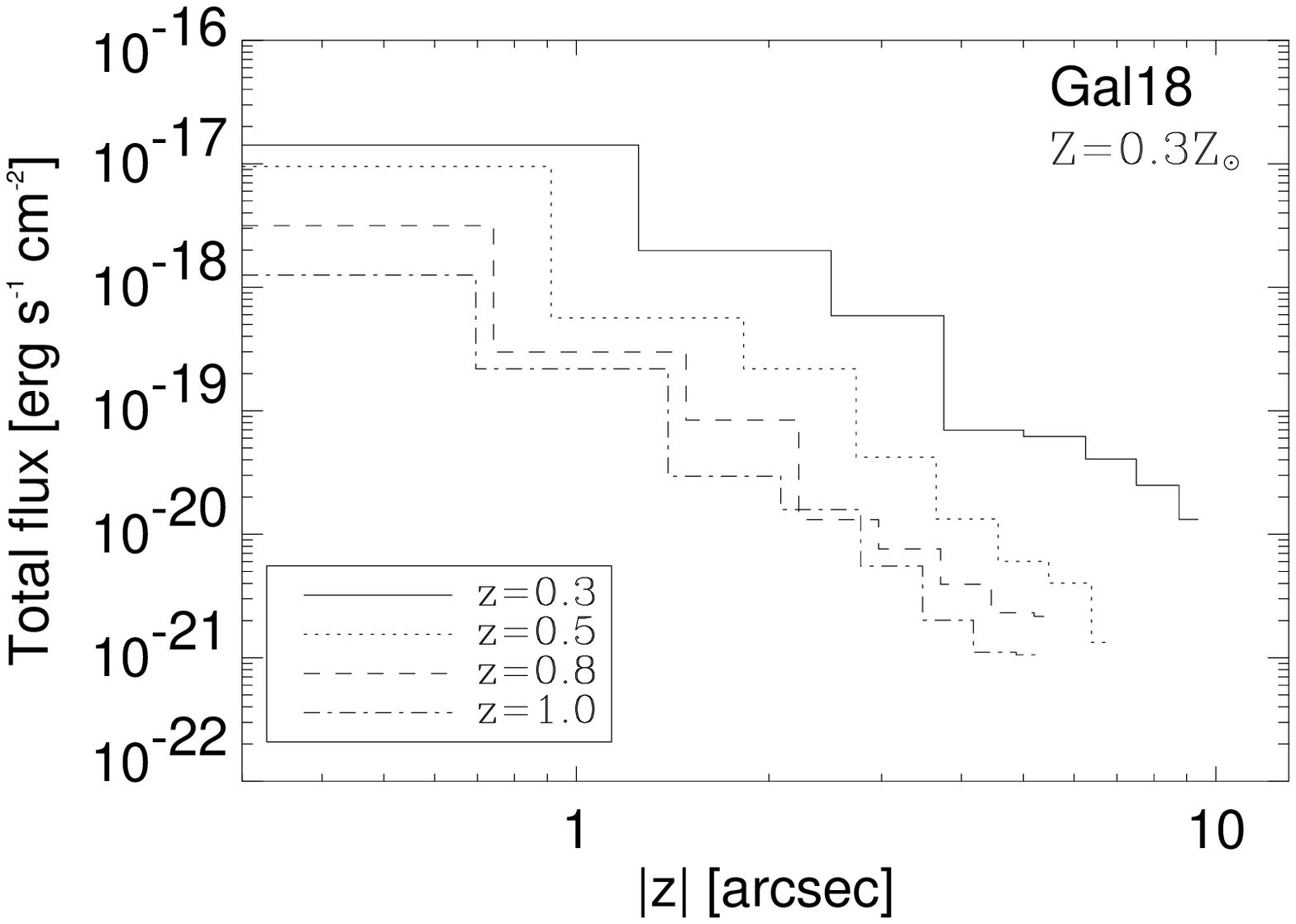}\hspace{0cm}}
\caption{{\it Top panel}: 0.3--2 keV surface brightness of the 
$Z=0.3$Z$_{\odot}$ galaxies inside 40 kpc wide slabs oriented along the disc, 
as a function of vertical distance $|z|$ to the disc. The galaxies are
viewed edge-on; for each vertical bin, the values above and below the disc 
have been added.
{\it Bottom panel}: Corresponding flux profiles 
(surface brightness integrated over
the sky area covered by the slabs), as a function of observed angular 
separation from the disc.}
\label{fig,surfbright}
\end{figure*}

(1) Due to the strong luminosity evolution of the $Z=0.3$Z$_{\odot}$ haloes
(Fig.~\ref{fig,lxz}), the 
optimum redshift for observing such haloes in terms of their 
surface brightness is not necessarily $z=0$ but is not well-defined either. 
If $L_X$ depends as strongly on redshift as for {\tt gal15}, the optimum 
redshifts could be $z\ga 0.5$, depending on the assembly history of the 
galaxy. 
For the $Z=0.0$ galaxies (not shown) there is a larger span between the flux
at different redshifts, with $z=0$ always being the optimum redshift.
The fluxes for haloes at $z>1$ are in all cases lower than those shown (this 
is also the case in the 0.1--3 keV band).

(2) To unambiguously detect the haloes in imaging observations, we estimate 
that a 
firm detection at least $\sim 10$ kpc above the disc is required in order to 
avoid confusion with disc emission. Taking {\tt gal15} as representative, it 
can be inferred from the bottom panel of Fig.~\ref{fig,surfbright}
that detection of haloes of MW-like galaxies at 
cosmological distances is clearly beyond the capabilities of current
instrumentation, despite the expected redshift evolution
of $L_X$. Even with the sub-arcsec spatial resolution of {\it Chandra} and 
point source sensitivities exceeding those of the {\it Chandra} Deep Fields 
by a factor of a few, a direct detection of halo emission would not be 
possible.

With an expected spatial resolution comparable to {\it XMM-Newton}, the 
next-generation mission {\it Constellation-X} (e.g.\ \citealt{whi99}) will 
not be able to image the haloes directly either. However, 
normal spiral discs in the nearby Universe typically exhibit two distinct 
spectral features, a $T\sim$ 0.2--0.3 keV thermal component from diffuse 
disc emission and a power-law from the point source population 
(e.g.\ \citealt{kun03}; \citealt{swa03}).
One may therefore rather seek to test the existence of the haloes and their 
predicted luminosity evolution by finding evidence for an additional 
low-temperature thermal component in the integrated spectra of disc galaxies.
To explore this possibility, we generated artificial source spectra for a 
{\it Constellation-X} observation of {\tt gal15} ($z=0.3$), using 
{\sc xspec} v11.0 along with the response matrix for the calorimeter 
detector\footnote{Available from 
{\tt http://constellation.gsfc.nasa.gov/docs/}}.
The adopted response assumes a four--spacecraft configuration with a total
effective area of 15,000 cm$^2$ at $E=1.25$ keV.  
The source was modelled as a combination of a $T=0.3$ keV thermal {\sc mekal}
plasma with $Z=$ Z$_{\odot}$ 
(representing disc diffuse emission), a $\Gamma=1.9$
power-law (representing point sources), and a {\sc mekal} plasma representing 
the hot halo, with each component normalized according to Figs.~\ref{fig,lxz} 
and \ref{fig,cdfn}.
We find that a 1 Ms observation will gather around 300 source counts over the 
full detector band (0.25-10 keV), of which only $\approx 10$ will 
originate in the
halo. These numbers will be down by an order of magnitude at $z=1$.
It thus seems impossible that {\it Constellation-X} should be able
to detect the haloes within reasonable exposure times. 
 
From the 0.1--3 keV profiles (not shown), one finds that extended energy 
coverage does result in larger observed flux, but the increase in emitted 
flux when including lower energies is nearly balanced by Galactic 
absorption. Depending on redshift, the increase in observed flux from 
0.3--2 keV to 0.1--3 keV is at most a factor of 1.5--2 for the
$Z=0.0$ galaxies and even less for $Z=0.3$Z$_{\odot}$, even for a relatively 
low column density $N_H$ as the one adopted. 
However, the Wide-Field Imager onboard the final configuration of
{\it XEUS}, with a spatial resolution
comparable to that of {\it Chandra} and an improved sensitivity particularly
at low energies
(an expected limiting point-source flux of 
$4\times 10^{-18}$ erg cm$^{-2}$ s$^{-1}$ for a 100 ks observation; 
\citealt{ble02}), should
observe $\sim 6000$ source counts in 1 Ms for {\tt gal15} at $z=0.3$, of which
$\sim 300$ would be from the halo. While this is not sufficient to warrant
the introduction of an additional thermal component in a spectral fit, the 
halo should be detectable as extended emission surrounding the disc in
soft-band images, after removal of disc and obvious point source emission.
We thus predict that {\it XEUS} should be able to single 
out halo emission of highly inclined disc galaxies out to $z\sim 0.3$, but it
will require exposure times comparable to those of the CDF.
At $z\approx 1$, about 120 halo counts out of 300 in total are expected. 
Although the ratio of halo to disc emission is significantly higher at this 
redshift, the lower flux and smaller spatial extent will make halo detection 
more difficult. 

The use of a different prescription for the disc SFR should
not impact on the halo detection prospects, but the inclination of the
galaxy will. 
The above results all apply to edge-on galaxies, i.e.\ a galaxies in which 
the polar axis is inclined by $i=90^{\circ}$ with respect to the 
line of sight. 
For a galaxy with smaller inclination, halo detection will be more difficult,
the effect in general depending on the 3-D distribution of hot halo gas. 
One may estimate the minimum galaxy inclination at which a halo is still 
detectable. As a first approximation, we can simply evaluate the sky area 
covered by the disc and assume that this area will effectively block all halo 
emission. This seems a reasonable assumption,
since all halo emission originating in front of the disc will be hard to 
distinguish from disc emission, while halo emission behind the disc will be 
subject to the same effect as well as suffer from severe absorption by 
neutral gas in the disc.
Although the method only takes into account the area blocked by the disc, 
not the fact that the halo itself will show a similar inclination, the
error implied should be negligible for large inclinations. Assuming a 
circular disc of radius 15 kpc and employing the $Z=0.3$Z$_{\odot}$ halo of 
{\tt gal15} at different redshifts, we estimate that 
at an inclination of $80^{\circ}$ around half of the halo emission will still
be detectable, whereas at $i=75^{\circ}$ this number is down to 
$\sim 30$ per cent. Adopting a tolerance level of 50 per cent, we thus
require the inclination 
to be $i\ga  80^{\circ}$, i.e.\ the disc should deviate less than $10^{\circ}$
from an edge-on appearance for the halo to remain detectable. This implies 
that $\sim 10$ per cent of all MW-like discs at a typical distance of 
$z=0.3$ should display a detectable halo in a 1 Ms {\it XEUS} observation.

The haloes of massive ($V_c\ga 300$ km s$^{-1}$) disc galaxies are predicted
to have X-ray luminosity and surface brightness at $z\simeq 0$ an order 
of magnitude larger than the haloes studied here (cf.\ Paper I). 
If the halo luminosities of such galaxies follow a similar behaviour with 
redshift to that shown in Fig.~\ref{fig,lxz}, and the galaxy shows an
inclination $i\ga 80^{\circ}$, these haloes should in fact be detectable to 
redshifts $z\sim 1$ and less inclined haloes to well beyond $z\approx 0.3$.
We can estimate the total surface density of detectable haloes on the sky by 
assuming that 10 per cent of all haloes around discs with $V_c \geq 220$ km
s$^{-1}$ can be detected to $z=0.3$. Using the velocity function of spirals
derived by \citet{gon00}, one arrives at a density of 
$\approx 10$ haloes deg$^{-2}$ in the adopted cosmology. 
Given our assumptions this would be a conservative lower limit, 
since some haloes beyond $z=0.3$ would also be 
detectable (those with $V_c$ well beyond 220 km s$^{-1}$), as would some 
at $z<0.3$ (where our inclination requirement, $i\ga 80^{\circ}$, is
severely relaxed).

\section{Summary and conclusions}\label{sec,sum}

From the X-ray properties of the hot gas haloes of two Milky Way--like disc 
galaxies 
extracted from a cosmological simulation, run both with primordial and 
intracluster chemical composition, we find that halo X-ray luminosities 
increase roughly 
an order of magnitude from $z=0$ to $z=1$, evolving more gently at higher
redshifts. There is good agreement between the redshift evolution of
$L_X\langle 1/T\rangle$ and that resulting from rough estimates of 
the disc accretion rate, as suggested by simple analytical considerations.
The logarithm of the disc accretion rate increases approximately linearly
with redshift out to at least $z\approx 1$, from a typical value of 
$\sim 0.5$ M$_{\odot}$ yr$^{-1}$ at $z=0$ to $\sim 3$ M$_{\odot}$ yr$^{-1}$
at $z=1$.

When added to a constant contribution
from diffuse gas in the disc and bulge along with a model predicting the
evolution of $L_X$ from X-ray binaries, we find that the halo $L_X$ of the 
galaxies is consistent with values derived for spirals in the 
CDF-N data. The luminosities of the $Z=0.3$Z$_{\odot}$ haloes 
suggest, however, that a constant halo metallicity of $0.3$Z$_{\odot}$ across 
the $z\sim $ 0--2 redshift range studied here
overpredicts the level of X-ray emission. Although this result is tentative
rather than conclusive and depends on the assumed evolution in disc star
formation rate, we speculate that halo 
metal abundances would be somewhat lower in the $z=0-2$ range than 
the typical metallicity of the intergalactic/intracluster medium (IGM) at 
$z\la 1$. Possible explanations
could include enriched supernova ejecta falling back on to the disc (galactic
chimneys), enriched gas being expelled out of the galactic gravitational 
potential
(possibly through supernova-driven superwinds during periods of strong star 
formation), or enriched gas being stripped from the haloes during 
galaxy-galaxy and galaxy-IGM interactions.

We have assessed the possibility for detecting haloes of MW-like galaxies and
shown that observations of such haloes at cosmological distances will have to
await a future generation of X-ray instrumentation. We find that 
{\it XEUS}
in its final configuration should be able to detect halo emission out to
at least $z\approx 0.3$ in a 1 Ms observation, and that the surface density
of haloes detectable in such observations should be $\ga 10$ deg$^{-2}$. 
In terms of 
surface brightness, the optimal redshifts for detecting haloes could
in fact be $0.5<z<1.0$, owing to their luminosity evolution.

It is worth emphasizing that these conclusions are based on a
particular numerical simulation with a certain set of input values; the aim
was to assess the amount of associated X-ray emission in this specific case
rather than to test the effects of varying the physical parameters
entering in the simulation (some effort was devoted to this in Paper I). 
Given the observational difficulty in directly
detecting the halo emission, the best way to test and improve on the 
predictions presented here in the immediate future may indeed be through 
additional simulations, perhaps coupled with 
advances in our still incomplete understanding of the formation of spiral 
discs. Higher numerical resolution and yet more realistic input physics 
with improved estimates for the important physical parameters such as baryon 
fraction and gas metallicity should refine the predictions presented here.
In particular, simulations of disc galaxy formation in which the gas chemical
evolution is self-consistently included may shed new light on the amount and 
physical state of hot halo gas around normal spiral discs
(Sommer-Larsen, Portinari \& Romeo 2003, in prep.). With the observed 
integrated X-ray emission from spirals acting as a basis, improved models for
the redshift evolution of disc X-ray luminosity would alternatively allow 
tighter constraints to be imposed on halo emission.

\section*{Acknowledgments}
We are grateful to the referee for prompt and constructive comments. 
This work was supported by 
the Danish Natural Science Research Council (SNF), 
the Danish Ground-Based Astronomical Instrument Centre (IJAF), 
the Carlsberg foundation, and by
Danmarks Grundforskningsfond through its support for the establishment of
the Theoretical Astrophysics Center.\\

\label{lastpage}

\end {document}